\documentclass[12pt]{amsart}

\usepackage{hyperref}

\newcommand {\ssecfont} {\normalfont}
\newcommand {\sssecfont} {\normalfont}

\newcommand {\bcdot} {\mathbin{\hbox{\raise.4ex\hbox{\bf.}}}} 

\newcommand {\ssbegin}[1]
 {\refstepcounter{subsection}
 \def \secno {\gdef \secno {}{\ssecfont
\thesubsection.\hskip 2ex}%
 }%
 \begin{#1}}

\newcommand {\sssbegin}[1]
 {\refstepcounter{subsubsection}
 \def \secno {\gdef \secno {}{\sssecfont
\thesubsubsection.\hskip 2ex}%
 }%
 \begin{#1}}

\begin{document}

\title[Queerification]{New simple Lie superalgebras as queerified associative algebras} 

\author{Dimitry Leites}

\address{New York University Abu Dhabi,
Division of Science and Mathematics, P.O. Box 129188, United Arab
Emirates\\ Department of Mathematics, Stockholm University, Roslagsv. 101,
Kr\"aftriket hus 6, SE-106 91 Stockholm, Sweden; dimleites@gmail.com}



\begin{abstract} 
Over $\mathbb{C}$, Montgomery superized Herstein's construction of simple Lie algebras from finite-dimensional associative algebras, found obstructions to the procedure and applied it to $\mathbb{Z}/2$-graded 
associative algebra of differential operators with polynomial coefficients. 

Since the 1990s, Vasiliev and Konstein with their co-authors constructed (via the Herstein--Montgomery method, having rediscovered it) simple Lie (super)algebras from the associative (super)algebra such as Vasiliev's higher spin algebras (a.k.a. algebras
of observables of the rational Calogero model) and algebras of symplectic reflections.

The ``queerification'' is another method for cooking  a simple Lie superalgebra from the simple associative (super)algebra. 
The (super)algebras of ``matrices of complex size'' and the above examples of associative (super)algebras can be ``queerified'' by adding new elements resembling Faddeev--Popov ghosts.

Conjectures: 1)  a ``queerified'' Hamiltonian describes  a version of the Calogero model with $1\vert 1$-dimensional time; 2) metabelean algebras and inhomogeneous subalgebras of Lie superalgebras naturally widen supersymmetries
in future theories; 
3) only certain graded-commutative algebras can imitate algebras of functions in  a reasonably rich non-commutative Geometry.
\end{abstract}



\maketitle

\markboth{\itshape Dimitry Leites} {{\itshape Simple Lie superalgebras as queerified associative algebras}}

\thispagestyle{empty}

\section{Preliminaries}

Simple Lie algebras and superalgebras carry important symmetries useful in innumerable applications. Here I describe a~new method producing many new simple Lie superalgebras.
Clearly, certain relatives of simple Lie (super)algebras are no less important in applications than the simple ones (e.g., Poisson Lie (super)algebra vs. Lie (super)algebra of Hamiltonian vector fields, affine Kac--Moody algebras vs. loop algebras, see an interesting discussion in \cite{BLS}). To describe such relatives is a~ task for future.

\subsection{From associative to Lie} Let $\mathbb{K}$ be an algebraically closed ground field of characteristic $p\neq 2$; unless otherwise stated, we consider $\mathbb{K}=\mathbb{C}$. For the case of $p= 2$ and finite-dimensional algebras, see \cite{BLLS}.

Let $A$ be any associative algebra, let $A^L$ be the Lie algebra whose space is $A$ but multiplication is given by the commutator $[a,b]:=ab-ba$. 

Similarly, if $A$ is a~ $\mathbb{Z}/2$-graded associative algebra, let $A^S$ be the Lie superalgebra with the multiplication given by the supercommutator. Let $\mathfrak{g}^{(1)}:=[\mathfrak{g},\mathfrak{g}]$ be the first derived Lie (super)algebra, a.k.a. (super)commutant, of the Lie (super)algebra $\mathfrak{g}$.

\sssbegin{Theorem}{\textup{(\cite{H})}}\label{ThH} Let $A$ be any simple finite-dimensional associative algebra with center $Z$ and not equal to it. Then, $L(A):=(A^L)^{(1)}/((A^L)^{(1)}\cap Z)$ is a~simple Lie algebra \emph{(unless $[A: Z]=4$ and $p=2$)}. \end{Theorem}


In \cite{Mon}, Montgomery generalized Herstein's result to $\mathbb{Z}/2$-graded simple associative algebras. (Actually, Montgomery considered so-called ``colored'' algebras, graded by any commutative group $G$, although it was already known, thanks to Scheunert, see \cite{Sch}, that there is a~natural equivalence between $G$-graded $G$-Lie algebras reducing the representatives of equivalence classes to either Lie algebras or to Lie superalgebras.)

Montgomery found out a~sufficient condition to the super version of Herstein's theorem~ \ref{ThH} and formulated it in the infinite-dimensional situation (in the finite-dimensional case this condition also works). 

\sssbegin{Theorem}{\textup{(\cite[Th.3.8]{Mon})}}\label{ThM} Let $A$ be a~$\mathbb{Z}/2$-graded simple associative algebra of characteristic $p\neq 2$ with supercenter $Z$ whose elements supercommute with any $a\in A$. Let the condition
\begin{equation}\label{Cond}
\text{if $u^2 \in Z$, then $u\in Z$ for any homogeneous $u\in A_{\bar 1}$}
\end{equation} 
hold. Then, 
\begin{equation}\label{SL(A)}
\text{SL}(A):=(A^S)^{(1)}/((A^S)^{(1)}\cap Z)
\end{equation}  
is a~simple Lie superalgebra.\end{Theorem}

\textbf{Examples}. If $\dim A<\infty$, then the simple Lie (super)algebras obtained by Herstein--Montgomery method are only those of series $\mathfrak{sl}$ or $\mathfrak{psl}$ (recall that $\mathfrak{psl}(pk):=\mathfrak{sl}(pk)/\mathbb{K} 1_{pk}$ if the characteristic $p$ of $\mathbb{K}$ is positive; or $\mathfrak{psl}(a|a):=\mathfrak{sl}(a|a)/\mathbb{K} 1_{2a}$ in the super case). If ${\dim A=\infty}$, the condition~ \eqref{Cond} was verified for algebras of observables in the rational Calogero model (a.k.a. symplectic reflection algebras) in \cite[Th.7.2]{KT1}, see also numerous simple examples in \cite{KS,KT2,KT3}.

\subsection{Central simple superalgebras} The associative algebra over a~field $\mathbb{K}$ is called \textit{central simple} if it has no proper ideals and its center is 1-dimensional over $\mathbb{K}$. Over any algebraically closed field $\mathbb{K}$ of characteristic $p\neq 2$, the only finite-dimensional central simple associative algebras are algebras $\text{Mat}(n)$ of $n\times n$ matrices. 

The super analog of this classification states that the only finite-dimensional central simple associative superalgebras are superalgebras $\text{Mat}(n|m)$ of supermatrices of size $n|m$ in the standard format, and 
\[
\text{Q}(n):=\{X\in \text{Mat}(n|n)\mid [X, J]=0\}, 
\]
where $J$ is an odd operator such that $J^2=a$ for an $a\in \mathbb{K}^\times$. If the characteristic of $\mathbb{K}$ is $p\neq 2$, we can assume that $a=-1$ and interpret the \textit{queer} algebra $\text{Q}(n)$ as preserving the complex structure given by an odd operator ~$J$. 

See also Finkelberg's reformulation of Wall's classification of central simple algebras in ``super'' terms, together with new results, in \cite[Ch. 7]{Lsos} or --- in English, but without the Brauer groups over $p$-adic fields and several other results due to Finkelberg --- Deligne's ``Notes on spinors'', \cite[pp. 99--135]{Del}. The answer over $\mathbb{C}$: any central simple superalgebra is either $\text{Mat}(a|b)$ or $\text{Q}(n)$. The classification over $\mathbb{R}$ explains Bott's periodicity; for more details, see \cite[Ch. 7]{Lsos}.

\subsection{Queerification in characteristic $p\neq 2$ (from \cite{BLLS})}\label{111} Let $A$ be an
associative algebra. The space of the associative algebra $\text{Q}(A)$ --- the \textit{associative queerification} of $A$ --- is $A \oplus\Pi(A)$, where
$\Pi$ is the change of parity functor, with the same multiplication in $A$ and the adjoint action of $A$ on $\Pi(A)$; let 
\[
\Pi(x)\Pi(y):= xy\text{~for any~~}x,y\in A.
\]
 Clearly, $\text{Q}(n)=\text{Q}(\text{Mat}(n))$. 

We will be mostly interested in the following \textit{\textbf{Lie}} version of queerification.

The space of the Lie superalgebra $\mathfrak{q}(A)$, 
called the
\textit{Lie queerification} of $A$, is $A^L\oplus\Pi(A)$, so $\mathfrak{q}(A)_{\bar 0}=A^L$ and $\mathfrak{q}(A)_{\bar 1}=\Pi(A)$, with the bracket given
by the following expressions and super anti-symmetry
\begin{equation}\label{commRel}
{}[x,y]:=xy-yx;\quad [x,\Pi(y)]:=\Pi(xy-yx);\quad [\Pi(x),\Pi(y)]:=
xy+yx \ \text{~for any~~}x,y\in A.
\end{equation}
The term ``queer", now conventional, is taken after the Lie
superalgebra $\mathfrak{q}(n):=\mathfrak{q}(\text{Mat}(n))$. (The associative superalgebra $\text{Q}(n)$ is an analog of $\text{Mat}(n)$; likewise, the Lie superalgebra $\mathfrak{q}(n)$ is an analog of $\mathfrak{gl}(n)$ for several reasons, for example, due to the role of these analogs in Schur's Lemma and in the classification of central simple superalgebras.) We express the elements
of the Lie superalgebra $\mathfrak{g}=\mathfrak{q}(n)$ by means of a~pair of matrices
\begin{equation}
\label{.1} (X,Y)\longleftrightarrow
\begin{pmatrix}X&Y\\Y&X
\end{pmatrix}
\in \mathfrak{gl}(n|n), \text{~~where $X,Y \in \text{Mat}(n)$}.
\end{equation}

For any associative $A$, we will similarly denote the elements of $\mathfrak{q}(A)$ by pairs $(X,Y)$, where $X,Y \in A$. The brackets between these elements are as follows:
\begin{equation}\label{.2}
\renewcommand{\arraystretch}{1.4}
\begin{array}{l}
{}[(X_1,0),(X_2,0)]:=([X_1,X_2],0),\quad
[(X,0),(0,Y)]:=(0,[X,Y]),\\
{}[(0,Y_1), (0,Y_2)]:=(Y_1Y_2+Y_2Y_1,0).
\end{array}\end{equation}

One can similarly queerify any associative \textbf{super}algebra $A$, see \cite{BLLS}. In particular, $\text{Q}(\text{Mat}(m|n))\simeq \text{Q}(m+n)$.

 Let $\mathfrak{sq}(n):=\mathfrak{q}(n)^{(1)}$ be the first derived algebra, the superalgebra of queertraceless matrices, where the \textit{queer
trace}, introduced in \cite{BL}, is given in notation of eq.~\eqref{.1} by the formula
\[
\text{qtr}:(X,Y) \mapsto \text{tr} Y.
\]
 The Lie superalgebras $\mathfrak{q}(n)$ and
$\mathfrak{sq}(n)$ are specifically ``super" analogs of the general Lie
algebra $\mathfrak{gl}(n)$ and its special (traceless) subalgebra $\mathfrak{sl}(n)$; we define their projectivizations to be $\mathfrak{pq}(n):=\mathfrak{q}(n)/\mathbb{K} 1_{2n}$ and
$\mathfrak{psq}(n):=\mathfrak{sq}(n)/\mathbb{K} 1_{2n}$.

\sssbegin{Theorem}\textup{(\cite{BLLS})} The only finite-dimensional simple Lie superalgebras related with queerification
in characteristic $p\ne2$ are $\mathfrak{psq}(n)$ for $n>2$. \end{Theorem}

\subsection{Two constructions of simple Lie (super)algebras from associative algebras}
These constructions are (1) the queerification and (2) the Herstein--Mont\-gomery construction (using a~$\mathbb{Z}/2$-grading). Both these constructions work in any characteristic of the ground field.

Recall that Lie superalgebras (actually, Lie super rings over~ $\mathbb{Z}$ and Lie superalgebras over finite fields) first appeared not as a~tool in high energy physics in 1974 or a~bit earlier (as many think), but in the 1940s with seedlings of idea in 1930s, in topology. Lately, simple (or close to simple, like $\mathfrak{gl}$ to $\mathfrak{psl}$) modular (i.e., over fields of characteristic $p>0$) Lie algebras and Lie superalgebras became a~topic of more general interest caused, e.g., by a~connection of the representations of quantum groups $U_q(\mathfrak{g})$, where $\mathfrak{g}$ is a~simple finite-dimensional Lie algebra over $\mathbb{C}$ and $q$ is the $p$th root of unity with the representations of an incarnation of $\mathfrak{g}$, or its super version, over the fields of characteristic $p$, see the book \cite{J} and a~lucid example in \cite{V}. 

\textbf{From Lie algebras to Lie superalgebras when $p=2$}. Interestingly, over an algebraically closed field of characteristic 2, \textbf{all} finite-dimensional simple Lie superalgebras are obtained by one of these two constructions
--- queerification and the Herstein--Montgomery method. Both constructions can be applied not only to associative algebras but also --- and this is vital --- to every simple Lie algebra, see \cite{BLLS}. 

\textbf{A~third construction when $p=3$ and $5$: from Lie algebras to Lie superalgebras with indecomposable integer Cartan matrix}. In characteristics $3$ and $5$, Kannan  suggested to construct Lie superalgebras from Lie algebras (only with indecomposable integer Cartan matrix so far) by using modern tools of mathematical physics such as Verlinde categories, see~ \cite{Kan}.

\section{Examples of associative algebras ready to be queerified}

In this section, I explicitly describe simple associative algebras $A$ related with various problems of interest in some areas of physics and mathematics, see \cite{Va,KV1,KV2,EG,Lo}, and of completely different nature, see \cite{Ha} and references therein. 

\ssbegin{Theorem} The queerifications $\mathfrak{q}(A)$ of the simple associative algebras $A$ described in this section are obtained by formulas~ \eqref{commRel}. Passing to the~subquotients we get simple Lie superalgebras. \end{Theorem}

\begin{proof} Directly follows from Theorem~\ref{ThM}.\end{proof}

\subsection{Recapitulation: a notion needed in Subsection~\ref{CalMoz}. Smash product of algebras (\cite{CM})} Recall that if $A$ is an algebra over a~field $\mathbb{K}$ and $\mathcal{H}$ is a~bialgebra with comultiplication denoted $\Delta(h)= \sum_{(h)} h_{(1)} \otimes h_{(2)}$ and counit $\varepsilon : \mathcal{H} \to \mathbb{K}$, then $A$ is called an $\mathcal{H}$-\textit{module algebra} if there exists a~map $m: A\otimes \mathcal{H} \to A$ such that
\[
\begin{array}{l}
\text{$A$ is an $\mathcal{H}$-module under $m$},\\
m(ab\otimes h) = \sum_{(h)} m(a\otimes h_{(1)}) \, m(b\otimes h_{(2)})\text{~ for any $a, b\in A$, $h\in \mathcal{H}$,}\\
m(1\otimes h) = 1_A\varepsilon(h)\text{~ for any $h\in \mathcal{H}$}.\\
\end{array}
\]
The \textit{smash product} $A\#\mathcal{H}$ of a~bialgebra $\mathcal{H}$ by an $\mathcal{H}$-module algebra $A$ is the space $A\otimes \mathcal{H}$ whose elements $a\otimes h$ are expressed as $a\#h$ with multiplication given by the next formula in which $\mu$ is the multiplication in $\mathcal{H}$ and $\cdot$ or juxtaposition denote the multiplication in $A$:
\[
(a\#g )(b\#h) := \sum_{(g)} a\cdot \big(m(g_{(1)}\otimes b)\# \mu(g_{(2)}\otimes h)\big) \text{~ for any $a, b\in A$, $g,h\in \mathcal{H}$}.
\]

\subsection{Symplectic reflection algebras (\cite{EG,KT1}) and algebras of observables in the rational Calogero model (\cite{KV3, KS,KT3}) a.k.a. higher-spin algebras (\cite{Va,KV2})}\label{CalMoz} Let $\Gamma\subset \text{Sp}(V)$ be a~finite group of automorphisms of a~vector space $V$ equipped with a~symplectic form. 
Recall that an element $\gamma \in \Gamma$ is said to be a~\textit{symplectic reflection} if $\text{rk}(\gamma- \text{id}) = 2$. 

Let $H_{\kappa}$ be a~ multi-parameter deformation of the smash product $S^{\bcdot}(V)\#\mathbb{C}[\Gamma]$ of the group algebra of $\Gamma$ by the polynomial algebra on $V$. The parameter ${\kappa}$ runs over points of $\mathbb{CP}^r$, where $r$ is the number of conjugacy classes of symplectic reflections in $ \Gamma$. The algebra $H_{\kappa}$ is called a~\textit{symplectic reflection algebra}. 

If $\Gamma$ is the Weyl group of a~root system in a~vector space $\mathfrak{h}$ and $V = \mathfrak{h}\oplus \mathfrak{h}^*$, then the algebras $H_{\kappa}$ are certain ``rational'' degenerations of the \textit{double affine Hecke algebra} introduced by Cherednik, see \cite{Ch}, and hence are sometimes called \textit{rational Cherednik algebras}.

A simplest example of $H_{\kappa}$ is the algebra $H_{1,c}$ with 
$c=\nu$ a.k.a. the \textit{algebra of observables in the $N$-particle rational Calogero model} with Hamiltonian 
\[
H_{\text{Cal}}:= -\frac12\sum_{1\leq i\leq N}\left(\frac{\partial^2}{\partial_{x_i}^2}-x_i^2
+\nu\sum_{j\neq i}\frac{2}{x_i-x_j}\frac{\partial}{\partial_{x_i}}\right),
\]
which can be expressed in terms of the Dunkl operators $D_i$ as the anticommutator of creation/annihilation operators:
\begin{equation}\label{Dunkl}
\begin{array}{l}
H_{\text{Cal}}:= -\frac12\sum_{1\leq i\leq N}\ [a_i^0, a_i^1]_+, \text{~~where}\\[2mm]
a_i^\alpha:=\frac{1}{\sqrt2}(x_i+(-1)^\alpha D_i),\text{~~where $\alpha\in\{0, 1\}$,}\\[2mm]
D_i:=\frac{\partial}{\partial_{x_i}}+\nu\sum_{j\neq i}\frac{2}{x_i-x_j}(1-K_{ij}),
\end{array}
\end{equation}
and where the $K_{ij}$ are the operators that permute indices:
\[
K_{ij}x_i = x_j K_{ij},\ \ K_{ij}x_k = x_k K_{ij} \text{~~for $k\neq i$, $k\neq j$.}
\]

The \textit{algebra of observables} is the associative algebra of polynomials in the $a_i^\alpha$ and $K_{ij}$.

Theorem \ref{ThLo} describes simple associative algebras $H_{{\kappa}}$ in an interesting particular case; for the general case, see Theorem \ref{ThLo1}.

\sssbegin{Theorem}{\textup{(\cite[Th.5.8.1, part 1)]{Lo})}}\label{ThLo} Let $\Gamma=S_n$.
The algebra $H_{1,c}$ is central simple if and only if $c\neq\frac{q}{m}$, where 
\begin{equation}\label{form}
\text{$q$ and $m$ are relatively prime integers and $1< m \leq n$.}
\end{equation} 
\end{Theorem}

\sssbegin{Theorem}{\textup{(\cite[Th.4.2.1]{Lo})}}\label{ThLo1} Let $\Gamma\subset \text{Sp}(V)$ be any finite group. Let $\overline{\mathbb{Q}}$ designate the field of algebraic numbers, let $r$ be the number of conjugacy classes of symplectic reflections in $ \Gamma$. There is a~finitely generated subgroup $\Lambda\subset \overline{\mathbb{Q}}^r$ such that the algebra $H_{1,c}$, where $c\in\overline{\mathbb{Q}}^r$, is central simple whenever
$\sum_{i=1}^r\lambda_ic_i\not \in \mathbb{Z}$ for all $(\lambda_i)_{i=1}^r\in\Lambda\setminus \{0\}$.
\end{Theorem}

\subsubsection{\textbf{Open questions}} 1) From the point of view of this paper, in order to get new simple Lie superalgebras, it is interesting to investigate the exceptions in the above Losev's theorems: the cases where the algebras $H_{1,c}$ have ideals. It is unclear if the quotient modulo such an ideal is simple, are the ideals nested, etc., cf. \cite{KT1,KT2,KT3}: in other words, what, if any exist, are the other central simple algebras obtained as quotients of $H_{1,c}$ or as its ideals? 

2) \textbf{Is there a~version of Calogero model with odd versions of Dunkl operators?}\label{Ra}
Let $P^-(x)$ be the polynomial algebra with generators $x_i$ for $i=1, \dots, n$ satisfying ${x_ix_j+x_jx_i=0}$ for $i\neq j$. This algebra is needed to construct (for a~long time elusive) odd versions of Dunkl operators, see~ \cite{Ra}. The center of $P^-(x)$ is generated by the $x_i^2$ for $i=1, \dots, n$, and hence Montgomery's condition \eqref{Cond} is violated.

I was unable to describe subquotients $A$ of $P^-(x)$ such that $\mathfrak{q}(A)$ is a~ simple Lie superalgebra, except for the known example: setting $x_i^2=a\in\mathbb{K}^\times$ we pass from $P^-(x)$ to its quotient, a~Clifford algebra $\text{Cliff}(n)$, which satisfies condition \eqref{Cond} and is ``naturally''
 $\mathbb{Z}/2$-graded by setting ${p_{\text{nat}}(x_i)={\bar 1}}$ for any $i$. The Herstein-Montgomery construction turns $\text{Cliff}(n)$ into the simple Lie superalgebra $\mathfrak{psq}(n)$, while the queerification of $\text{Cliff}(n)$ considered as a~superalgebra yields $\mathfrak{psq}(2n)$ after the passage to a~ subquotient.

\subsection{The Lie algebra of ``complex size'' matrices $\mathfrak{gl}(\lambda)$ and its generalizations}\label{sllambda} Let $\mathfrak{g}$ be any simple finite-dimensional Lie algebra over $\mathbb{C}$, let $\rho$ be the half-sum of positive roots of $\mathfrak{g}$, let $\mu$ be the highest weight of an irreducible $\mathfrak{g}$-module with highest weight vector $v^\mu$, let the $C_i$ be the Casimirs, i.e., the generators of the center of $U(\mathfrak{g})$. Set
\begin{equation}\label{U_c}
U_c:=U(\mathfrak{g})/(C_i-c_i)_{i=1}^{\text{rk} \mathfrak{g}}, \text{~~where $c_i=C_i|_{v^{\mu+\rho}}\in \mathbb{C}$}.
\end{equation} 
For almost all values of $c=(c_1,\dots, c_{\text{rk}\mathfrak{g}})$, the algebra $U_c$ is central simple, cf. the example below. 

In the particular case of $\mathfrak{g}=\mathfrak{sl}(2)$ and $c_1=\lambda^2-1$, where $\lambda=\mu+1\in \mathbb{C}$ 
for the highest weight $\mu$ of an irreducible $\mathfrak{sl}(2)$-module, the algebra $U_\lambda:=U(\mathfrak{g})/(C_1-c_1)$ has an ideal
\[I_\lambda=\begin{cases}0&\text{if $\lambda\not\in \mathbb{Z}\setminus\{0\}$},\\
\text{of finite codimension}&\text{otherwise}.\\
\end{cases}
\]
If $\lambda=n\in \mathbb{Z}\setminus\{0\}$, then $U_n$ has an ideal $I_n$ of finite codimension
and $U_n/I_n \simeq \text{Mat}(|n|)$, hence the name of $\mathfrak{gl}(\lambda)$ for any $\lambda\in\mathbb{C}$, see \cite{Fei}. If $\lambda\not\in \mathbb{Z}\setminus\{0\}$, then $U_\lambda$ and its limit as $\lambda\to\infty\in\mathbb{C}\mathbb{P}^1$, see \cite{GL}, are central simple as well as $U_n/I_n$. 

The Lie algebras $(U_c)^L$ are ``quantized'' versions of the algebras of functions on the orbits of the co-adjoint representation of simple finite-dimensional Lie algebras over $\mathbb{C}$, see \cite{Kon}, considered with the Poisson bracket.

Observe that, having defined the parity of a root as the parity of the corresponding root vector, and $\rho$ as a~half sum of positive even roots minus a~half sum of positive odd roots, it is possible to apply definition \eqref{U_c} to Lie superalgebras $\mathfrak{g}=\mathfrak{osp}(1|2n)$, and get central simple superalgebras fit to be queerified. (Recall that one can queerify not only associative algebras, but associative superalgebras as well.) 

For any other finite-dimensional Lie superalgebra $\mathfrak{g}(A)$ with indecomposable Cartan matrix $A$ or the simple subquotient of $\mathfrak{g}(A)$, the definition~ \eqref{U_c}, where $i$ runs~ 1 through $\text{rk}\,\mathfrak{g}$, is inapplicable because the center of $U(\mathfrak{g}(A))$ is finitely generated only for $\mathfrak{g}(A)=\mathfrak{osp}(1|2n)$. But this does not prevent us from considering the infinite set of generators $C_i$ of the center of $U(\mathfrak{g})$ and use generalized formula (7) in which $c_i=C_i|_{v^{\mu+\rho}}$ for all $i$.

Observe that there are simple finite-dimensional Lie superalgebra $\mathfrak{g}$, without Cartan matrix, but still with sufficiently big center of $U(\mathfrak{g})$, e.g., the quotient of the first derived of the Poisson superalgebra $\mathfrak{po}(0|2n)$ modulo center, see \cite{LS}.

Observe that for simple finite-dimensional Lie superalgebras, the algebra of rational functions in Casimirs is finitely generated by the first $\text{rk}\ \mathfrak{g}$ Casimirs (as was first observed in some cases by Berezin; for a~complete description in the case where $\mathfrak{g}$, or its double extension like $\mathfrak{gl}(n|n)$, see \cite{BLS}, has a~ Cartan matrix, see~ \cite{Sg}).

In the cases where the center of $U(\mathfrak{g})$ is trivial, i.e., consists only of constants (e.g., if $\mathfrak{g}=\mathfrak{spe}(n)$, the supertraceless subalgebra of the periplectic Lie superalgebra $\mathfrak{pe}(n)$ which preserves a~non-degenerate \textbf{odd} anti-symmetric bilinear form, i.e., $\mathfrak{pe}(n)$ is the linear part of the antibracket superalgebra), one can apply Serganova's construction from~\cite{Ser}, i.e., replace the algebra $U(\mathfrak{g})$ with $\overline U(\mathfrak{g}):=U(\mathfrak{g})/\mathfrak{r}(U(\mathfrak{g}))$ whose center is sufficiently big, where $\mathfrak{r}(A)$ is the radical of the algebra $A$. Serganova's idea is applicable to other simple Lie superalgebras
$\mathfrak{g}$ with the trivial center of $U(\mathfrak{g})$ (and with non-trivial centers as well). Conjecturally, one can then replace in eq.~\eqref{U_c} the algebra $U(\mathfrak{g})$ with $\overline U(\mathfrak{g})$ thus enlarging the set of examples of superalgebras of ``complex size matrices''.

For various applications of (super)algebras of ``complex size matrices'', see \cite{LS1, Sg1, Sg2}, \cite{GL1} and references therein.

\subsection{Simple reduced $C^\ast$-algebras of groups} Certain huge central simple algebras $A$ are needed to produce $C^\ast$-simple groups  important in operator theory.  I'll quote from a~very lucid review MR2303514 (2008a:22004) by Bachir Bekka:

``Important examples of $C^\ast$-algebras arise from unitary representations of locally compact groups. If $G$ is such a~group and $(\pi , H)$ a~unitary representation of $G$, then $\pi $ “extends” to a~representation of the convolution algebra $L^1(G)$ and the closure in the operator norm of $\pi (L^1(G))$ in $B(H)$ [the latter being the collection of bounded linear operators on $H$] is a~$C^\ast$-algebra. Applied to the regular representation, which is defined on $L^2(G)$ through the action of $G$ by left translations, this construction yields the so-called \textit{reduced $C^\ast$-algebra} of $G$, denoted by $C_r^\ast(G)$. The group $G$ is said to be $C^\ast$-\textit{simple} if $C_r^\ast(G)$ has no nontrivial closed two-sided ideals.
The paper \cite{Ha} is a~comprehensive and lively exposition on the question: Which groups are $C^\ast$-simple?" 

In \cite{Ha}, de la Harpe gave a~rather long list of such groups, hence of central simple algebras $C_r^\ast(G)$ pertaining to our study in this paper as associative algebras that can be queerified into a~simple Lie superalgebra. For a~description of centers in several such associative algebras, see~ \cite{Los}. 

I am not sure if the really huge Lie superalgebras $\mathfrak{q}(A)$ thus obtained are of interest. However, encouraging are examples of certain $\mathbb{Z}$-graded Lie (super)algebras of exponential growth that  were ignored until recently as too huge to be of use but lately found various applications (by Borcherds, Gritsenko and Nikulin).

\section{Dynamics with $1|1$-dimensional time. How to queerify Hamiltonians?}
The rectifiability of the vector fields on supermanifolds with local coordinates $u_1, \dots, u_m$ (even) and $\xi_1, \dots, \xi_n$ (odd) states that the non-vanishing at some point vector field $D$ can be reduced in a~ vicinity of this point to the shape $\partial_{u_1}$ if the field is even or $\partial_{\xi_1}+\xi_1\partial_{u_1}$ if $D$ is odd and $D^2\neq 0$, see \cite{Sh}. (If $D^2=0$ for an odd non-vanishing at some point vector field $D$, then locally $D$ can be reduced to the shape $\partial_{\xi_1}$, which yields ``not interesting" Time and dynamics.) Let $\{-,-\}$ be either the Poisson bracket or the anti-bracket. Then, the most natural from super point of view mechanics should be given by the equation with $1|1$-dimensional time with coordinates $t$ (even) and $\tau$ (odd):
\begin{equation}\label{sqrt}
D_\tau(f)=\{f, \mathcal{H}\},\text{~~where $D_\tau:=\partial_{\tau}+\tau\partial_{t}$ and $p(\{-,-\})+p(\mathcal{H})=1$}.
\end{equation}
Since $D_\tau^2=\partial_t$, the equation \eqref{sqrt} should describe more fine structure of the dynamics than the conventional equation
\begin{equation}\label{usu}
\dot f=\{f, H\}, 
\end{equation}
used, e.g., in the classical description of the spinning top in \cite{BM}. (Clearly, $H:=\frac12\{\mathcal{H}, \mathcal{H}\}$.) To find out what extra phenomena are described by eq.~\eqref{sqrt} as compared with eq.~\eqref{usu} is an \textbf{open problem}. Both these equations describe dynamics of the same supermanifold. (M.~Vasiliev observed that the systems \eqref{sqrt} and \eqref{usu} are the simplest examples of \lq\lq unfolded systems", see \cite{MV}.)

Introducing analogs of Faddeev--Popov ghosts (elements of $\Pi(A)$) we ``double'' the supermanifold whose algebra of functions is $A$. \textbf{Conjecturally}, to get the dynamics corresponding to the queerified algebra of observables of the rational Calogero model we should ``queerify" its Hamiltonian by introducing ghosts, and, moreover, consider the $1|1$-dimensional time. 

\section{On great expectations} 

\subsection{On Differential Geometry with metabelean algebra of functions and Volichenko algebras replacing Lie superalgebras} Recall that a~ \textit{Volichenko algebra} is any inhomogeneous (with respect to parity) subalgebra of a~Lie superalgebra. In \cite{LSe1, LSe2}, the first and very interesting examples of \textit{simple} Volichenko algebras (defined as those without two-sided proper ideals) were first given, and the idea to consider the metabelean algebras (satisfying the identity $[a, [b,c]]=0$ for any three of its elements, where $[a,b]:=ab-ba$) as algebras of functions was elaborated. Since any metabelean algebra can be realized as a~subalgebra, perhaps inhomogeneous, i.e., not a~sub\textbf{super}algebra, of an enveloping supercommutative superalgebra, see \cite{KR}, there are two natural categories of supermanifolds: 

(A) In this category, currently considered by all researchers, only parity-preserving automorphisms of superalgebras are allowed under the universal belief that ``only them are physically meaningful"; this belief is tacitly or explicitly assumed starting with \cite{BeL}. (I think this belief will not last and will disperse, like certain ``no-go" theorems and general myths of the ``pre-super" time, i.e., before \cite{WZ}, that \lq\lq it is impossible to mix bosons with fermions".)

(B) In the other category, any automorphisms of superalgebras are allowed as was initially suggested in \cite{L0}, see also \cite{Dj, Ba}. (Observe that there are not that many ``additional" automorphisms, e.g., for the Grassmann algebra $G(n)$, the parity-preserving automorphisms depend on $n$ functional parameters, whereas any, not necessarily parity-preserving, automorphism depends on $n+1$ functional parameters.)

In the category (B), the spaces are not $\mathbb{Z}/2$-graded, but filtered: each space~ $V$ contains a~ subspace~ $V_{\bar 0}$ called \textit{even}, the \textit{odd} part of $V$ is not necessarily a~direct summand, but the quotient modulo~ $V_{\bar 0}$. For every Volichenko algebra~ $V$, its even part $V_{\bar 0}$ is a~Lie algebra, but the complementary space is not a~direct summand.

Iyer proved that \textit{Volichenko algebras} can serve as algebras of derivations of metabelean algebras, playing for them the role of Lie (super)algebras of derivations of (super)commutative algebras of functions, see \cite{Iy}. 

The idea to consider elements of metabelean algebras as functions, and Iyer's result, widened supersymmetry in a~natural way and not far: metabelean algebras (of functions) live inside supercommutative superalgebras while their algebras of derivations (analogs of Lie algebras) live inside Lie superalgebras. Interestingly, the changes of coordinates not preserving parity do allow integration on any supermanifold with any even number of odd coordinates, see \cite{Lcr}.

Strangely, this --- natural --- generalization of supersymmetry did not yet draw any interest of physicists and the fruits of this generalization are unknown so far. Interestingly, whereas the even wave functions corresponding to bosons are invariant under arbitrary automorphisms of algebras considered in the category (B), the notion of ``odd'' function is not invariant, so the answer to the question ``does a~given function correspond to a~fermion or not?'' depends on our choice of ``coordinates" (parity) defined modulo the even subspace.

\subsection{Dreams of graded-commuta\-tive Differential Geometry} In \cite{L0}, I introduced the ``something'' (superscheme) on which the elements of the Grassmann algebra play the role of functions, thus proving Berezin's conjecture ``There exists an analog of Calculus in which the elements of the Grassmann algebra play the role of functions" in the algebraic setting. The paper \cite{BeL} gave an equivalent (for proof, see \cite[Section~4.8]{Mo}) construction (though in the language of charts and atlases) in the smooth case (supermanifolds).

A local question ``Is there an analog of formula \eqref{Dunkl} with odd Dunkl operators?'' leads to a~most tempting global conjecture generalizing Berezin's conjecture by replacing the Grassmann algebra with $P^-(x)$, see Section~\ref{Ra}. 
The algebra $P^-(x)$ appeared in \cite{Schw} as a~conjectural quantum analog of the algebra of functions. However, until now nobody was able to construct an analog of Calculus or Differential Geometry where elements of $P^-(x)$ play the role of functions, unless one imposes conditions $x_i^2=0$ for all $i$ and gets the well-known now ``super Geometry", see \cite[Ch.~1]{Del}, \cite{Lsos}. In \cite{BM}, Berezin and Marinov observed that Schwinger did not explicitly state what the $x_i^2$ are equal to, evading this question. 


\textbf{Conjecture}. In a~reasonably rich generalization of Geometry (with integration and differential equations), only certain $G$-graded-commutative algebras can play the role of algebras of ``functions''. 

For example, after Morier-Genoud and Ovsienko understood how to endow the Clifford algebra with a~grading which turns it into a~ graded-commutative algebra, see \cite{MGO}, they were able to construct a~ determinant of the $n\times n$ matrices with values in the Clifford algebra, although --- mysteriously --- not all dimensions $n$ of the ``space'' permit to define this determinant, see \cite{COP}. Is it possible to endow the Weyl algebra, and the tensor product of Weyl and Clifford algebras, with a~$G$-grading turning them into $G$-graded-commutative algebras or the role of functions in this future geometry can be played only by the usual functions with values in the Clifford algebra? 

Observe, for the sake of completeness, that earlier than Scheunert, in an unpublished manuscript, Nekludova proved an analog of Scheunert's theorem for $G$-graded-commutative algebras (``under a~natural equivalence there are either commutative algebras or supercommutative superalgebras"); Molotkov sharpened both Scheunert's and Nekludova's results from $G$ finite (in the original Nekludova's claim) to any \textit{finitely generated} commutative group $G$ (as is tacitly assumed in \cite{Sch}, although just ``any'' is written), see \cite[Ch.8]{Lsos}.


\subsection*{Acknowledgements} I was supported by the grant AD 065 NYUAD.

\end{document}